\documentclass[doublecol]{epl2} 

\usepackage{amsmath}
\usepackage{bbm}
\usepackage[skip=6pt]{caption}

\title{Drift of interfaces in forced stably-stratified turbulence and the role of vertically-sheared helical structures}

\author{Niccolò Cocciaglia\inst{1}
\thanks{Email: \email{niccolo.cocciaglia@roma2.infn.it}},
Luca Biferale\inst{1}, 
Fabio Bonaccorso\inst{1} \and 
Alessandra S. Lanotte\inst{2}
}

\institute{                    
  \inst{1} Department of Physics \& INFN, University of Rome ``Tor Vergata'', 
  Via della Ricerca Scientifica 1, 00133, Rome, Italy\\
  \inst{2} CNR NANOTEC \& INFN Sezione di Lecce, 
  Via Monteroni, 73100 Lecce, Italy
}
\pacs{nn.mm.xx}{First pacs description}

\abstract{
Experimental investigations of forced stably stratified turbulence (SST) have shown that the step-like density profile, made of well-mixed density layers and sharp interfaces alternating along the gravity direction, undergo a slow coarsening dynamics with either decay or merging of interfaces. In this Letter, we focus on the coarsening dynamics phenomenon, by means of Direct Numerical Simulations of forced SST at moderate resolutions, and very long temporal integration. We show that the vertical drift and merging of interfaces is associated to the emergence of spatially-uniform, vertically-sheared helical structures that break the mirror-symmetry of the system. When these are absent, interfaces decay is observed instead. 
A dynamical correspondence between helicity dissipation rate by buoyancy effects and the vertical buoyancy flux allows to establish a (causal) connection between the chiral structures and the vertical movement of interfaces leading to merging.
}

\begin{document}

\maketitle

\section{Introduction}

Turbulent flows with stable density stratification display several specific features when compared to typical homogeneous isotropic turbulence: 
for instance, gravity-induced dispersive waves propagate local perturbations throughout the domain, and wave-breaking events (caused by either shear or convective instabilities) produce local mixing of momentum and density \cite{davidson_book, caulfield2021review, couchman2023mixing}. 
An external stirring mechanism, perturbing the hydrostatic linear density profile, feeds the instabilities responsible for the formation of anisotropic, flattened structures \cite{howland2020mixing}. 
Several experimental investigations \cite{park1994turbulent, holford1999mixing, thorpe2016} and recent numerical work \cite{petropoulos2023disruption, cocciaglia2025longtime} investigated the formation of a fully-layered state, made of stacked quasi-homogeneous density layers, spanning the whole domain in the horizontal directions, connected by thin interfaces with strong density variation. 
Long duration of the experiment and a strong-enough stratification strength (typically measured with the Froude number $Fr$) seems to be required to achieve this state, and for this reason most numerical simulations failed to observe a fully-formed layered structure. 
The layered state undergoes a coarsening dynamics \cite{park1994turbulent} with very long inter-event times (also thousands of large-eddy turnover times \cite{cocciaglia2025longtime}), and the type of coarsening falls into two categories, decay or merging. 
The former, at moderate Froude numbers $Fr=0.22$, sees a density interface being eroded and disappearing with little to absent vertical displacement, whereas the latter, at smaller Froude $Fr=0.076$, manifests itself with a pair of interfaces moving vertically towards each other until merging. 
In a companion paper \cite{cocciaglia2025longtime} we investigated the long-time behavior of forced stably-stratified turbulence, showing the occurrence of both coarsening mechanisms when stratification strength is varied. Remarkably, it was observed that while during the interface decays the kinetic energy remained statistically stationary, during interface merging events the kinetic energy exhibited strong and bursty fluctuations mostly associated to larger horizontal velocities.

In this Letter, we further investigate the interfaces evolution, and try to relate it to the kinetic energy excursions observed in the low-$Fr$ run. 
To this end, we perform a Craya-Herring decomposition of the velocity field and show that the dominating kinetic contribution comes from vertically-sheared horizontal flows, also known as shear modes. We notice how such type of flows is associated to the presence of large helicoidal structures that break the mirror symmetry of the system. Finally, we highlight a connection between the interface merging mechanism and the helicity decay due to buoyancy.
\begin{figure*}[t]
    \centering
    \includegraphics[width=\textwidth]{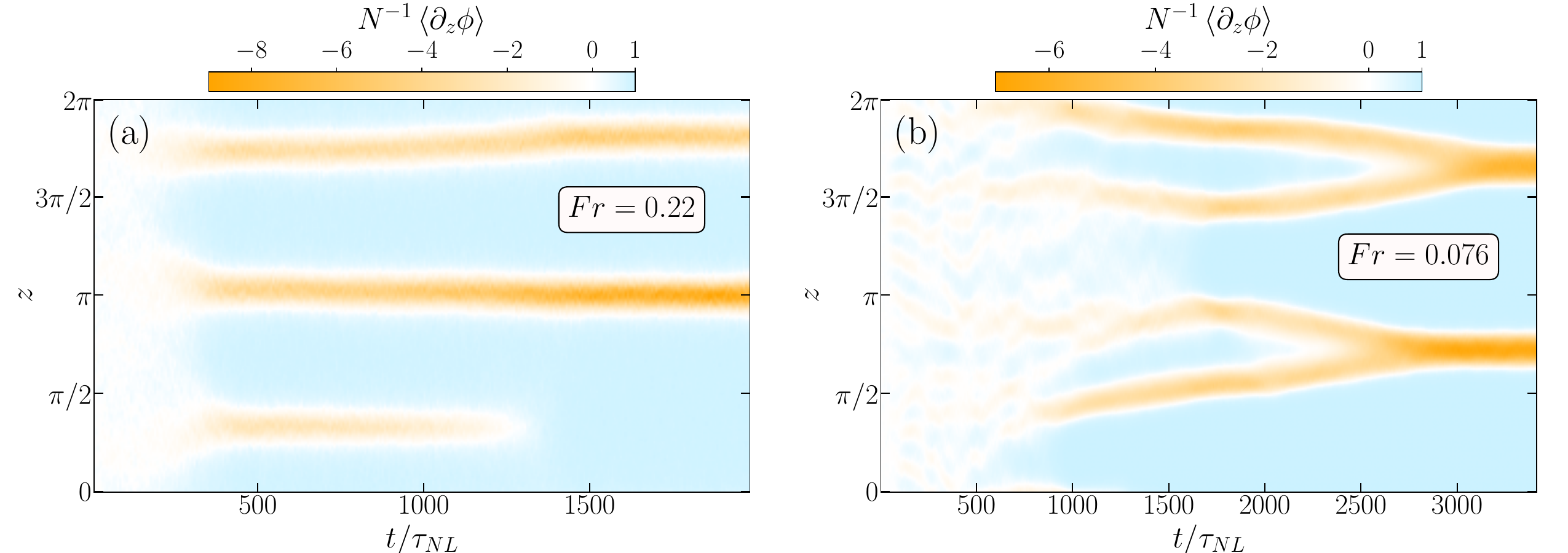}
    \caption{Evolution of the plane-averaged vertical gradient of the scalar field, rescaled by the Brunt-Vais\"al\"a frequency, for (a) $Fr=0.22$ ($N=6$) and (b) $Fr=0.076$ ($N=18$). Blue-shaded regions are layers, orange-shaded ones are interfaces.}
    \label{fig:gradphi_colormaps}
\end{figure*}


\section{Equations and numerics}
The forced stably-stratified flow is simulated from the Navier-Stokes equation in the Boussinesq approximation, with a pseudo-spectral simulation in a triperiodic cubic box of side $2\pi$. The coupled equations for the incompressible velocity $\boldsymbol{u}$ and the (rescaled) density fluctuation $\phi$ are:
\begin{align}
\begin{split}
    \partial_t \boldsymbol{u} + (\boldsymbol{u}\cdot \boldsymbol{\nabla}) \boldsymbol{u} &= -N \phi \hat{z} - \boldsymbol{\nabla}P/\rho_0 + \boldsymbol{\mathcal{D}}_\nu + \boldsymbol{\mathcal{D}}_\alpha + \boldsymbol{f}\,
    \label{eq:u_evol}
\end{split}\\[.1cm]
    \partial_t \phi + (\boldsymbol{u} \cdot \boldsymbol{\nabla}) \phi &= N \boldsymbol{u} \cdot \hat{z} + \boldsymbol{\mathcal{D}}_\kappa
    . \label{eq:phi_evol}
\end{align}
The scalar field $\phi=(N \rho_0/g) \delta\rho$ is related to the density fluctuation $\delta \rho$ superimposed to the linear density profile $\rho_l=\rho_0 - \sigma z$, and $N=\sqrt{(g/\rho_0) \partial_z\rho_l}$ is the Brunt-Vais\"al\"a frequency. An isotropic, stochastic forcing $\boldsymbol{f}$ excites the velocity field in the Fourier shell $4 \leq k_f \leq 6$, while 
the small-scale viscous and diffusive terms are 
$\boldsymbol{\mathcal{D}}_\nu\equiv - \nu_q (-\Delta)^q \boldsymbol{u}$ and $\boldsymbol{\mathcal{D}}_\kappa\equiv- \kappa_q (-\Delta)^q \phi$, respectively, and $q=2$. Last, a damping term acts in Fourier space as $\tilde{\boldsymbol{\mathcal{D}}}_\alpha = \sum_{k_1<|\boldsymbol{k}|<k_2} [-\alpha \tilde{\boldsymbol{u}}(\boldsymbol{k})/|\boldsymbol{k}|^2]$, removing kinetic energy with an inverse Laplacian at the largest scales ($k_1=0.5$, $k_2=2.5$), with damping coefficient $\alpha$. 
In this letter, we discuss results of two simulations with different stratification strengths, quantified by the Froude number defined as $Fr=U_{rms}/(N\, L_f)$, where the rms velocity averaged over the volume is $U_{rms} = \sqrt{\langle u_x^2 \rangle_V + \langle u_y^2 \rangle_V + \langle u_z^2 \rangle_V}$ and the characteristic scale is the forcing one $L_f = 2\pi/5$, with $k=5$ being the centroid of the forced shell.  Choosing Brunt-Vais\"al\"a frequency as $N=\{6,18\}$, we get the corresponding $Fr=\{0.22, 0.076\}$. 
Simulation times are rescaled with the nonlinear large-eddy turnover time $\tau_{NL} = L_f / U_{rms}$.
The initial condition for both runs are $\boldsymbol{u}=0$ and $\phi=0$, namely a fluid at rest in hydrostatic balance (linear density profile).
Volume-averaged quantities are denoted by $\langle \cdot \rangle_V$, whereas in the rest of the paper averages computed on horizontal $xy$ planes, frequently used in the study, will be represented as $\langle \cdot \rangle$. Quantities with tilde ($\tilde{A}$) denote Fourier representations.
More details on the simulations can be found in \cite{cocciaglia2025longtime}.


\section{Interface decay and interface merging in numerical simulations}

\begin{figure}[t]
    \centering
    \includegraphics[width=\linewidth]{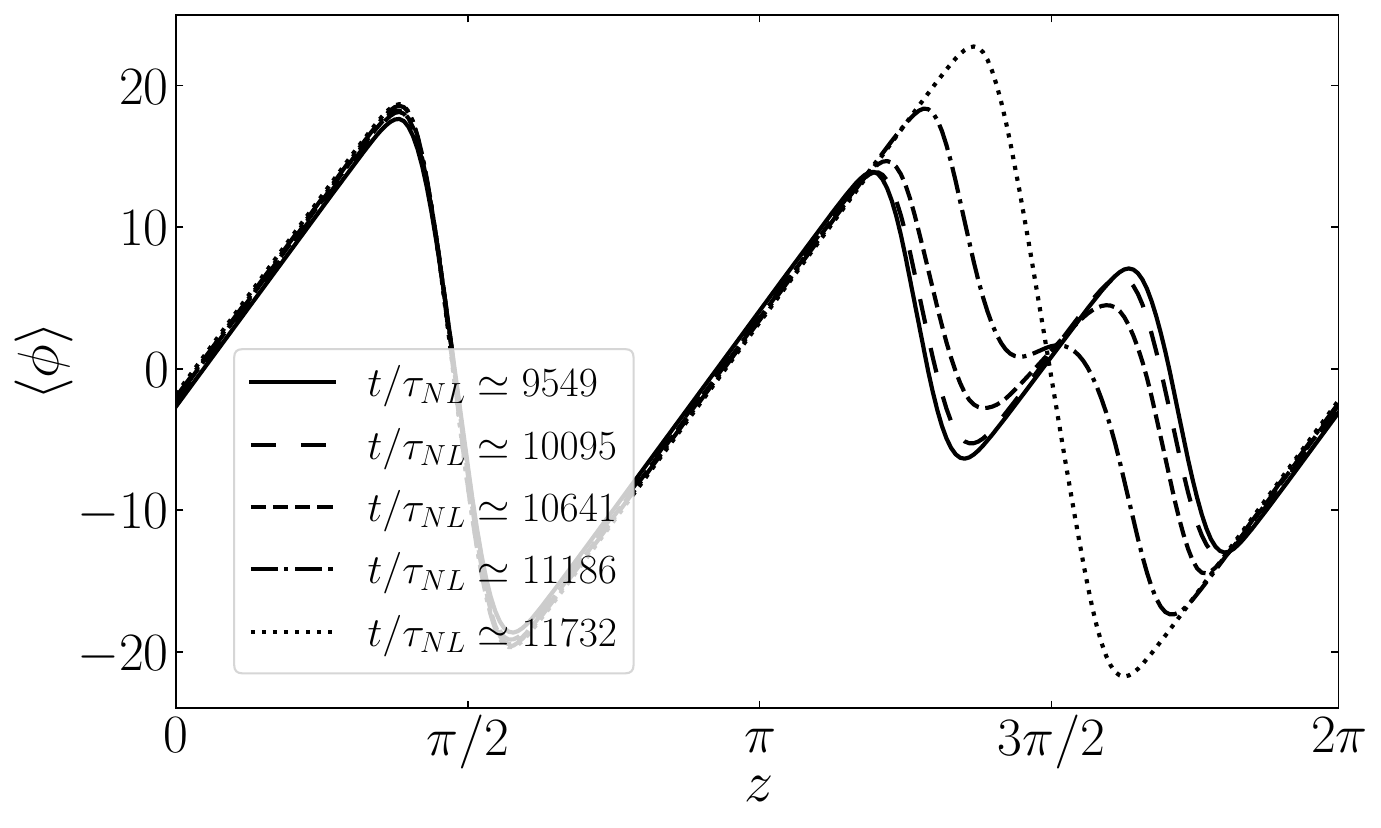}
    \caption{Five time snapshots of the ramp-cliff structure formed by the vertical profile of $\phi$, for $Fr=0.076$. On the right a merging event is observed, as two interfaces (negative slope) come close together and the layer between (positive slope) disappears. Data from \cite{cocciaglia2025longtime}.}
    \label{fig:interfacemerging_SofiaN18}
\end{figure}
 We start by providing a visual representation of the two different coarsening mechanisms that reduce the number of layers and interfaces, thus increasing the layers' thickness.
As shown in \cite{park1994turbulent,cocciaglia2025longtime}, by exploring turbulence-stratification parameters space, it emerges that the stratification strength determines the turbulent {\it equilibrium-state} of the system, i.e. whether a fully-layered state will form and how it will later evolve. This was confirmed in our DNS at fixed Reynolds numbers, where the emergence of the characteristic step-like density profile from the initial linear one was observed for $N=6$ and $N=18$, while it was absent for $N=3$ (not shown).
In terms of the rescaled density fluctuation $\phi$ the vertical profile is a ramp-cliff structure, as it is called in the context of passive scalar advection with a superimposed mean gradient \cite{pumir1994numerical, buaria2021smallscale}. 

In Fig.~\ref{fig:gradphi_colormaps} we show, by means of the plane-averaged vertical gradient of $\phi$, the occurrence of the two coarsening mechanisms: interface decay for $Fr=0.22$ (panel a) and interface merging for $Fr=0.076$ (panel b). 
As can be observed, decay is characterized by very little vertical movement of the overall structure, 
and the disappearance of an interface corresponds to full mixing between two adjacent layers. 
On the other hand, interface merging is associated to an overall vertical drift of existing interfaces. When pairs of interfaces come close they tend to coalesce into a single, stronger interface. 
An alternative view of merging in terms of the aforementioned ramp-cliff structure of $\langle \phi \rangle$ is represented in Fig.~\ref{fig:interfacemerging_SofiaN18}.

Some caution is required when making comparisons between numerical simulations in periodic domains and experiments in closed tanks. Indeed, the clear transition from decay to merging at varying $Fr$, observed numerically, is not seen in experiments, where both mechanisms were observed to happen at fixed $Re,Fr$, and during different stages of the experiment. 
It is important to note that the experimental coarsening is highly influenced by the no-flux boundary condition that a finite domain enforces. To get a unified understanding of these different observations, we remark that results of our tri-periodic simulations should be seen as if they took place in the bulk of the tank, far from the top surface and bottom boundaries.


\section{Velocity decomposition and the growth of shear energy}

\begin{figure}[t]
    \centering
    \includegraphics[width=\linewidth]{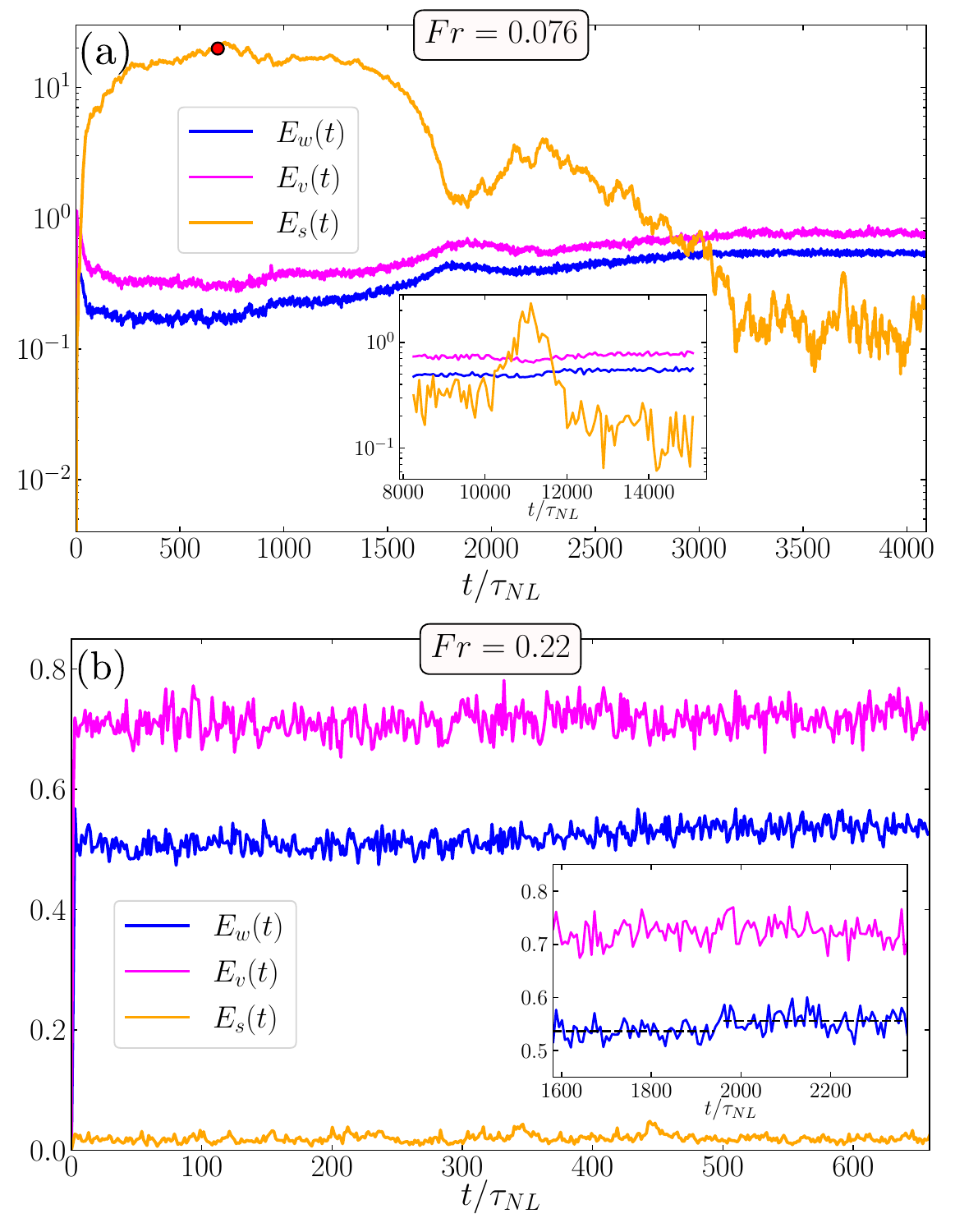}
    \caption{Time evolution of the wave, vortical and shear energies for (a) $Fr=0.076$ and (b) $Fr=0.22$. Main panels show the dynamics starting from a state of rest, while insets display a single, isolated coarsening event, extrapolated from a previous study \cite{cocciaglia2025longtime}: interface merging in (a) and interface decay in (b). Dashed lines in the inset of (b) are best fits performed before and after the interface decay. In panel (a) a $xz$ slice of the shear velocity modulus at $t/\tau_{NL} \simeq 680$ (red circle in main panel) is shown in the top right corner.}
    \label{fig:enekin_decomp}
\end{figure}

In \cite{cocciaglia2025longtime}, it is highlighted that surges of kinetic energy followed by a return to a stationary value occur during the vertical movement of interfaces. 
It has been long known \cite{smith2002generation, laval2003forced, lindborg2006cascade} that forced stratified flows experience an unbounded growth of vertically-sheared horizontal velocities, also known as `shear modes', that end up dominating over the contributions of waves and vortices. 
To assess the relation between kinetic energy build-up and interface vertical movement, we perform a Craya-Herring decomposition \cite{riley1981direct, laval2003forced} of the velocity field into its wave ($w$), vortical ($v$), and shear ($s$) components. 
In Fourier space the spectral velocity $\tilde{\boldsymbol{u}}(\boldsymbol{k},t)$ is constrained by incompressibility to evolve on a plane perpendicular to $\boldsymbol{k}$. 
On such plane two perpendicular directions are selected:
\begin{equation}
    \boldsymbol{e}_v = (\boldsymbol{k} \times \hat{\boldsymbol{z}})\, /\, \left|\boldsymbol{k} \times \hat{\boldsymbol{z}}\right|
    \label{eq:dir_vortical}
\end{equation}
\begin{equation}
    \boldsymbol{e}_w = [\boldsymbol{k} \times(\boldsymbol{k} \times \hat{\boldsymbol{z}})]\, /\, \left|\boldsymbol{k} \times(\boldsymbol{k} \times \hat{\boldsymbol{z}})\right|
    \label{eq:dir_wave}
\end{equation}
associated respectively to the vortical and wave components of the spectral velocity, $\tilde{\boldsymbol{u}}_v(\boldsymbol{k})$ and $\tilde{\boldsymbol{u}}_w(\boldsymbol{k})$. 
The former restricted to the horizontal plane is the one determining the vertical vorticity; the latter, perpendicular to the previous one, has non-zero vertical component and describes the vertical motion induced by internal waves. 
The decomposition \eqref{eq:dir_vortical}-\eqref{eq:dir_wave} breaks down when $\boldsymbol{k} \parallel \hat{\boldsymbol{z}}$: the corresponding modes with $k_\perp=\sqrt{k^2_x+k^2_y}=0$ are the shear modes $\tilde{\boldsymbol{u}}_s(\boldsymbol{k})= \tilde{\boldsymbol{u}}_s(k_\perp=0,k_z)$. 
The isotropic spectra associated to each component are computed as:
\begin{equation}
    \tilde{E}_c(k,t) = \sum_{k\leq |\boldsymbol{k}|<k+1} |\tilde{\boldsymbol{u}}_c(\boldsymbol{k},t)|^2/2\ , \quad c=w,v,s,
    \label{eq:spectra_wvs}
\end{equation}
from which the mean energies are:
\begin{equation}
    E_c(t) = \sum_k \tilde{E}_c(k,t)\ .
    \label{eq:enekin_wvs}
\end{equation}
We show first in Fig.~\ref{fig:enekin_decomp} the time evolution of the wave, vortical and shear energies from the initial state of rest, for $Fr=0.076$ in panel (a) and $Fr=0.22$ in panel (b). 
As expected, the shear modes are responsible for the growth of kinetic energy at smaller $Fr$. At their maximum, they are almost two orders of magnitude larger than wave and vortical contributions, and it takes thousands of turnover times for the shear energy to become subdominant with respect to the other contributions. 
For $Fr=0.076$, a vertical slice of the shear velocity modulus, computed close to the kinetic energy peak (red circle in main panel) is shown on the top right corner of Fig.~\ref{fig:enekin_decomp}(a): by definition it has no horizontal variability, and more energetic (lighter) bands alternate with less intense (darker) ones. 
On the other hand, at larger $Fr$ (panel b) the shear modes play a negligible role on the energetics, and only a slight increase of the wave energy is detected in correspondence to the formation of density layers, around $t/\tau_{NL} \simeq 400$. 
In the insets of Fig.~\ref{fig:enekin_decomp} we use data from~\cite{cocciaglia2025longtime}, and show the effect of a coarsening event (e.g. interface merging for $Fr=0.076$ and interface decay for $Fr=0.22$p) on the kinetic energy components. 
A slight increase of wave energy is detected in both cases, while for lower $Fr$ we see an excursion of shear energy signaling the merging event. 

\begin{figure}[t]
    \centering
    \includegraphics[width=\linewidth]{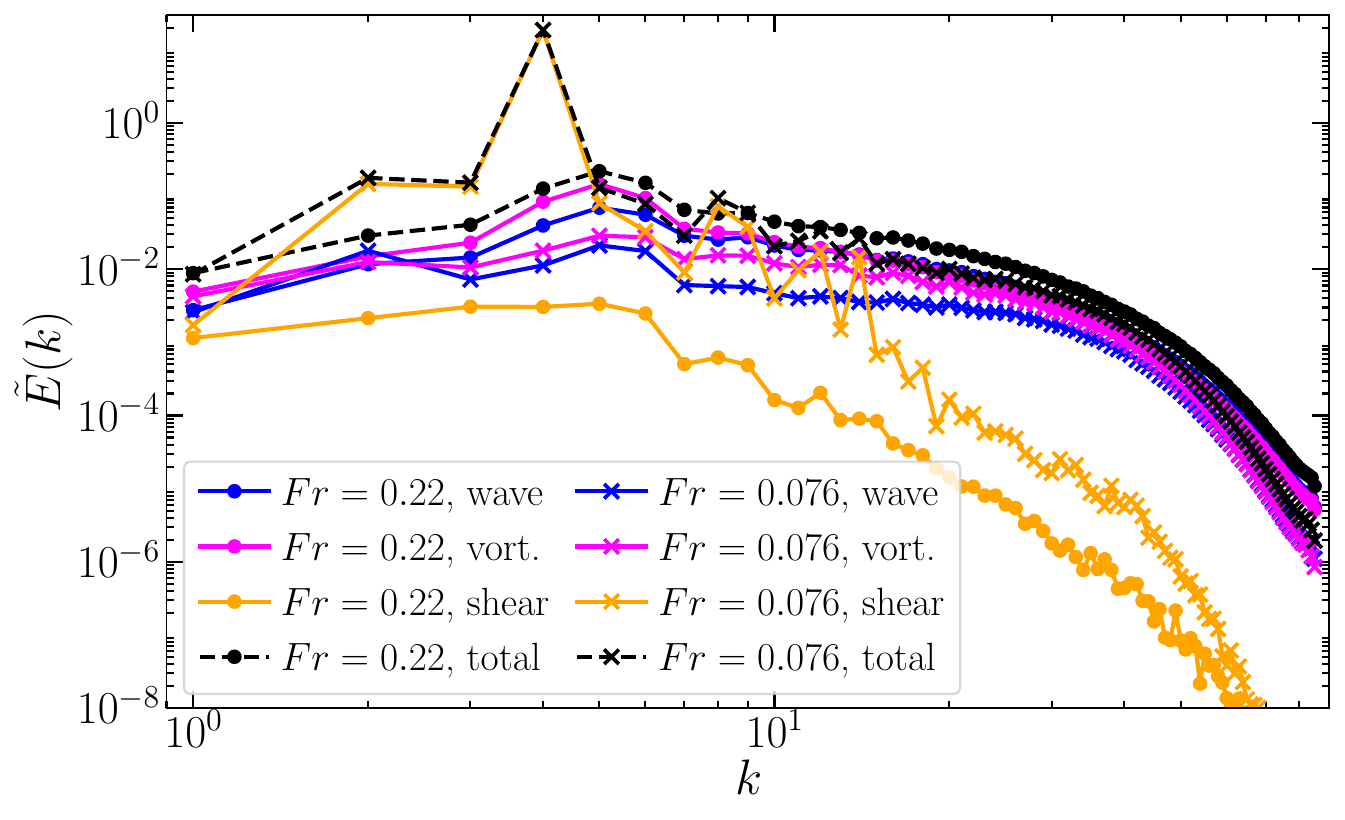}
    \caption{Isotropic spectra of the decomposed and total kinetic energies at $t/\tau_{NL} \simeq 650$, where for $Fr=0.22$ (circles) the total kinetic energy is stationary whereas for $Fr=0.076$ (crosses) it is close to the peak of its excursion. To achieve some smoothing, a $10$-samples time average is performed.}
    \label{fig:kin_spectra_decomp}
\end{figure}
Looking in Fig.~\ref{fig:kin_spectra_decomp} at the spectral densities of the three modal contributions (computed in correspondence to the states of largest kinetic energy), we notice that for both cases, the wave and vortical spectra peak at the forcing scales $4 \leq k_f \leq 6$. Differently, the shear energy, while being subdominant at all scales for $Fr=0.22$, it becomes up to 3 orders of magnitude larger than the other components at its peak for $Fr=0.076$. 
Moreover, a very fast decay of shear energy across the inertial range is noticed, steeper than a $k^{-4}$ power law. 


\section{Interface merging: the role of helicoidal flow}

Regarding the process of coarsening by merging, we seek for the cause producing enhanced vertical movement of the flat density structures and the coalescence of close interfaces, at sufficiently strong stratification. We have seen that vertically-sheared horizontal flows (i.e. shear modes) are completely dominant on the other kinetic contributions when vertical rearrangement of the density profile is taking place. 
In other words, large and purely horizontal (and horizontally coherent) velocities seem to promote vertical advection of the scalar field, but how?

\begin{figure}
    \centering
    \includegraphics[width=\linewidth]{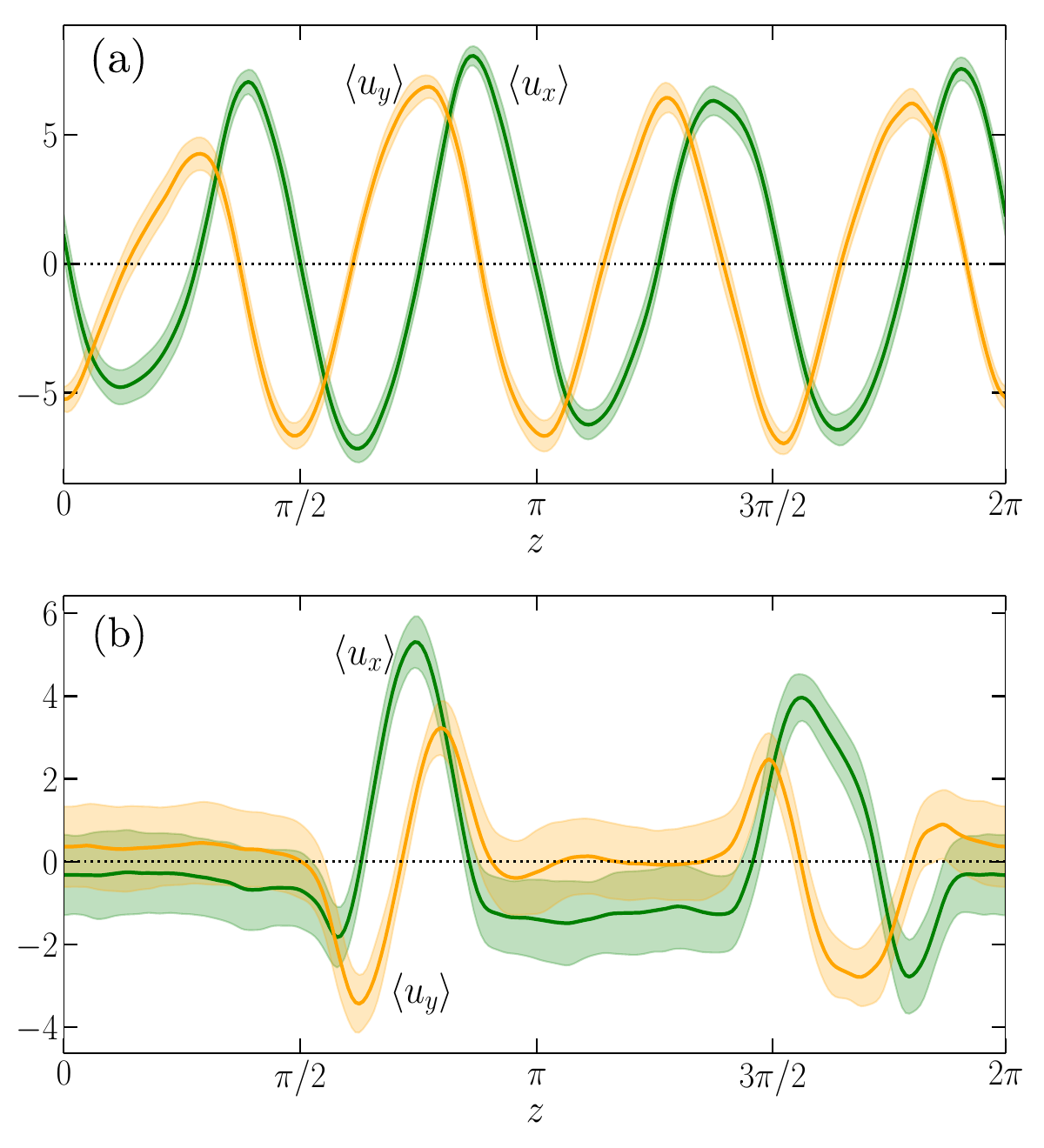}
    \caption{Vertical profiles, for $Fr=0.076$, of the horizontal velocity components $\langle u_x \rangle$ (in green) and $\langle u_y \rangle$ (in orange), at different instants: (a) at the peak of shear energy, $t/\tau_{NL}\simeq 750$; (b) when two parallel merging events are occurring, $t/\tau_{NL} \simeq 2350$. Shaded areas are the errors quantified by standard deviations from the plane-sampling.}
    \label{fig:horizvel_profiles_N18}
\end{figure}
\begin{figure}
    \centering
    \includegraphics[width=\linewidth]{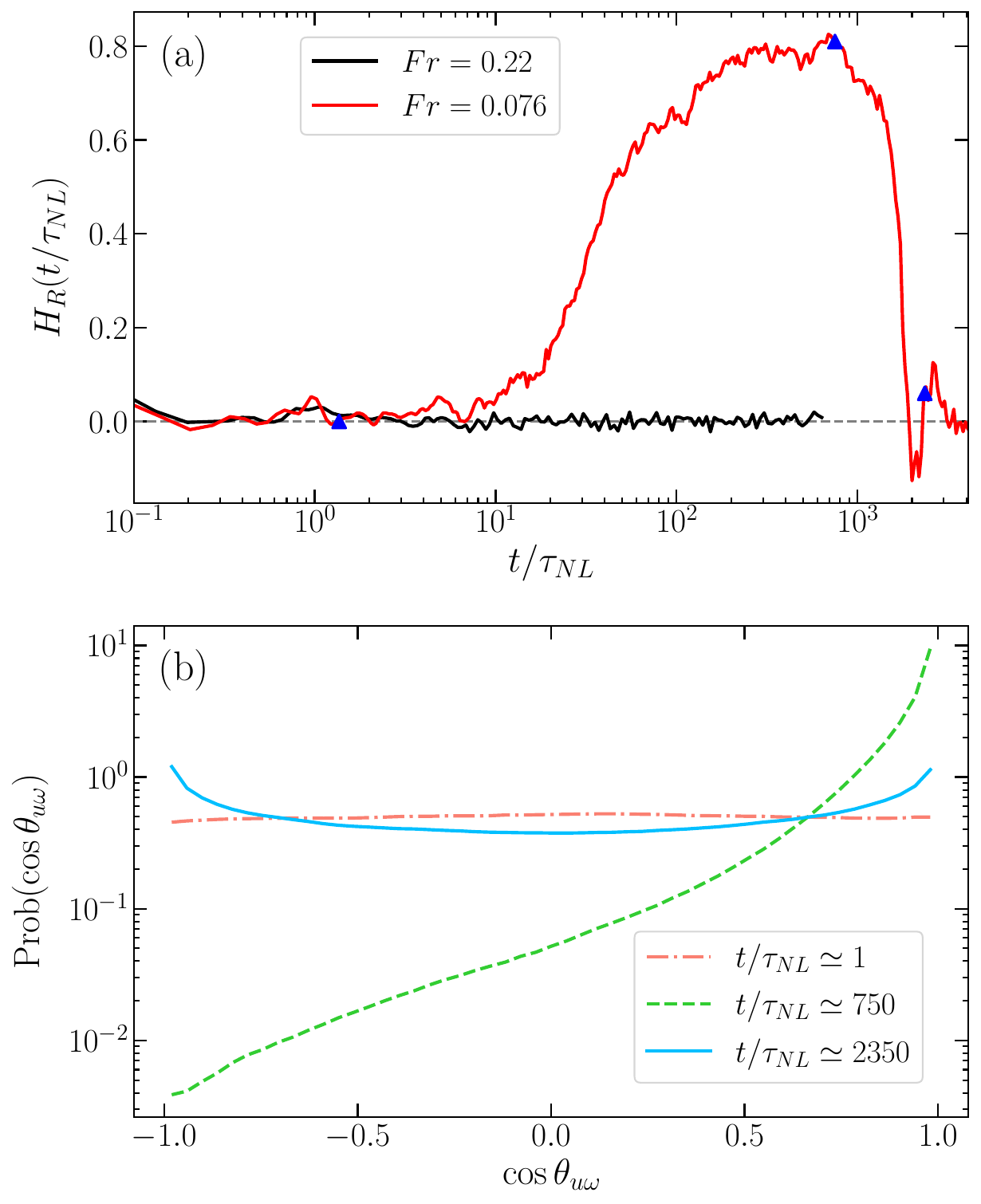}
    \caption{(a) Evolution of the relative helicity $H_R$ for the two stratification strengths, in logarithmic time axis. (b) Probability distributions of $\cos\theta_{u\omega} = (\boldsymbol{u}\cdot\boldsymbol{\omega}) / ( |\boldsymbol{u}|\, |\boldsymbol{\omega}| )$, where $\theta_{u\omega}$ is the angle between velocity and vorticity vectors, sampled for the $Fr=0.076$ run at three time instants specified in the legend,  and also indicated by blue triangles in panel (a). Two out of three time instants are the same as Fig.~\ref{fig:horizvel_profiles_N18}.}
    \label{fig:helicity_vs_time_and_pdf}
\end{figure}
It is interesting to notice that a specific pattern of the horizontal velocity emerges when shear modes are dominant and density structures show vertical drift: as it can be appreciated in Fig.~\ref{fig:horizvel_profiles_N18}(a), a very clean sine-cosine vertical profile is observed, with wavelength $\lambda = 2\pi/4$ as expected from the peak of shear energy at $k=4$ displayed in Fig.~\ref{fig:kin_spectra_decomp}. 
This result, along with the constraints that periodic boundaries put on $u_z$, tells that the plane-averaged velocity components become, in first approximation,
\begin{equation}
    \left(\langle u_x \rangle, \langle u_y \rangle, \langle u_z \rangle \right) = \left(A\, \sin(4 z+\psi), A \cos(4 z + \psi), 0\right) 
    \label{eq:circ_pol_wave}
\end{equation}
at the peak of kinetic energy (where the phase $\psi$ is arbitrary due to periodicity along $z$). 
Shear Fourier modes take the shape, in real space, of a domain-wide, vertically-oriented circularly-polarized wave, frozen in time and spiraling -- in our case -- like a left-handed screw. 
In order to asses whether the surges of shear energy are always associated to this kind of helical flow, we report in Fig.~\ref{fig:horizvel_profiles_N18}(b) the horizontal velocity profiles at a later instant, approximately where two simultaneous merging events occur: in the flat regions interested by the coarsening a right-handed (at lower $z$) and left-handed (at higher $z$) screws are observed. 

The breaking of mirror-symmetry in fluid systems is quantified by non-zero hydrodynamic helicity, defined locally as $h = \boldsymbol{u} \cdot \boldsymbol{\omega}$, where $\boldsymbol{\omega} = \boldsymbol{\nabla} \times \boldsymbol{u}$ is the vorticity. 
The degree of chirality can be measured by the (volume-averaged) relative helicity:
\begin{equation}
    H_R(t) =  \frac{\langle \boldsymbol{u} \cdot \boldsymbol{\omega} \rangle_V}{\sqrt{\langle |\boldsymbol{u}|^2 \rangle_V\ \langle |\boldsymbol{\omega}|^2 \rangle_V}}\ ,
    \label{eq:relative_helicity}
\end{equation}
with the property that $H_R \in [-1,1]$. 
Maximally-helical flows have $H_R = \pm 1$, and indeed the velocity field \eqref{eq:circ_pol_wave} belongs to the subclass of ABC flows \cite{dombre1986ABC} with $B=C=0$, that have maximum helicity (in modulus) by construction. 
Computing the relative helicity for the two runs we see in Fig.~\ref{fig:helicity_vs_time_and_pdf}(a) that, while no net helicity is detected at $Fr=0.22$, a large positive value ($H_R > 0.8$) is associated to the coherent polarized structure of the velocity field seen at lower $Fr$. 
Notice that when the two structures with opposite chirality form (Fig.~\ref{fig:horizvel_profiles_N18}(b)) the overall relative helicity oscillates between positive and negative values, depending on which helicoidal part gives the largest contribution at any given time. 
{\color{black}In support of the fact that the chirality of the ensuing structures is somewhat arbitrary, we add that the $Fr=0.076$ simulation studied in \cite{cocciaglia2025longtime} produced an early, domain-wide helical structure of negative helicity ($H_R(t/\tau_{NL}) \simeq -0.6$ at its minimum).}
We further quantify the alignment of velocity and vorticity vectors by computing the probability distribution of $\cos\theta_{u\omega} = (\boldsymbol{u}\cdot\boldsymbol{\omega}) / ( |\boldsymbol{u}|\, |\boldsymbol{\omega}| )$, the cosine of the angle between velocity and vorticity fields. 
As evident from Fig.~\ref{fig:helicity_vs_time_and_pdf}(b), at early times ($t \sim \tau_{NL}$) the distribution is uniform denoting no preferential alignment. 
When the shear energy is maximal ($t \sim 750\, \tau_{NL}$) the distribution peaks at $\theta_{u\omega}=0$, whereas when the two helices with opposite polarization are present ($t \sim 2350\, \tau_{NL}$) we observe alignment and anti-alignment -- $\theta_{u\omega}=0,\pi$ respectively -- with equal probability.

\begin{figure}
    \centering
    \includegraphics[width=\linewidth]{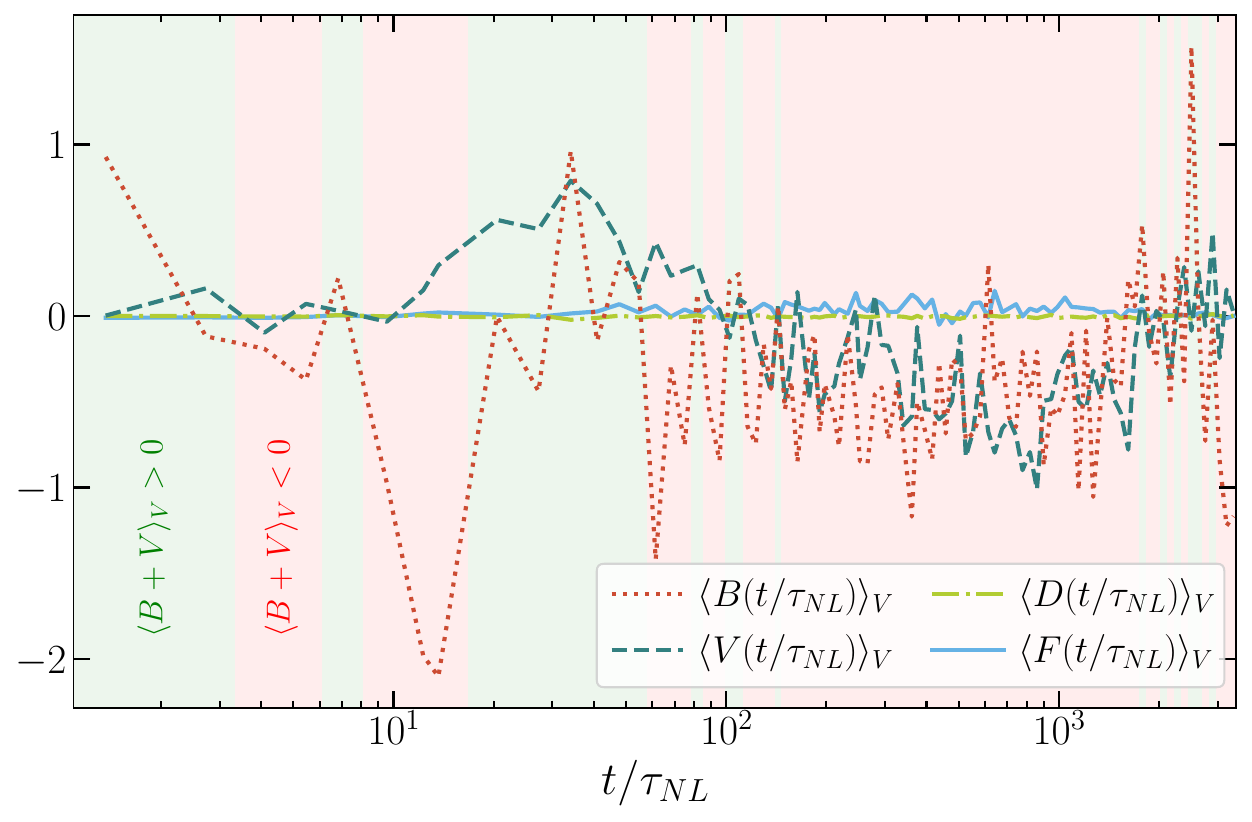}
    \caption{Dynamics of the volume-averaged values of the terms producing (or destroying) point-wise helicity, due to: buoyancy $B$, viscosity $V$, large-scale damping $D$ and external forcing $F$ (see eq.~\eqref{eq:helicity-dynamics}), with logarithmic time axis. Green bands signal where the sum of buoyancy and viscous terms is positive, red bands where it is negative. Data refer to the $Fr=0.076$ case.}
    \label{fig:helicitybalance_volumeavg_vs_time}
\end{figure}

In a Boussinesq fluid, the volume-averaged helicity $\langle h \rangle_V = \langle \boldsymbol{u} \cdot \boldsymbol{\omega} \rangle_V$ satisfies the budget equation below  \cite{lilly1986helicity, hide1989superhelicity}:
\begin{equation}
    \partial_t \langle h \rangle_V = 2\, \langle B+V+D+F\rangle_V\ .
    \label{eq:helicity-dynamics}
\end{equation}
Here the four terms on the r.h.s. are the contribution to global production or dissipation of helicity due to -- respectively -- buoyancy, viscosity, large-scale damping and external forcing. 
They read as:
\begin{align}
    \langle B \rangle_V &= -N\langle \phi \omega_z \rangle_V\ ,
    \label{eq:helprodrate_by_buoyancy} \\
    \langle V \rangle_V &= \nu_2\, \langle \boldsymbol{\omega}\cdot \Delta (\boldsymbol{\nabla}\times\boldsymbol{\omega}) \rangle_V\ ,
    \label{eq:helprodrate_by_viscosity} \\
    \langle D \rangle_V &= - \langle h \ast G \rangle_V \ ,
    \label{eq:helprodrate_by_damping} \\
    \langle F \rangle_V &= \langle \boldsymbol{f} \cdot \boldsymbol{\omega} \rangle_V \ .
    \label{eq:helprodrate_by_forcing}
\end{align}
in which $\ast$ denotes convolution, and $G$ is the kernel function whose Fourier representation is $\tilde{G}(\boldsymbol{k}) = \alpha |\boldsymbol{k}|^{-2}\, \mathbbm{1}_{k_1<|\boldsymbol{k}|<k_2}$. 
For completeness' sake, we write the evolution equation of the local helicity $h(\boldsymbol{r},t)$ for SST in appendix A of the Supplementary Material (SM). Note that the viscous term $\langle V \rangle_V$ in \eqref{eq:helprodrate_by_viscosity} involves order-2 hyper-dissipation: the equivalent terms for standard dissipation can be recovered removing the laplacian and changing sign. 
Figure~\ref{fig:helicitybalance_volumeavg_vs_time} shows the time evolution of the terms \eqref{eq:helprodrate_by_buoyancy}-\eqref{eq:helprodrate_by_forcing} determining the mean-helicity budget. 
The forcing and damping terms are clearly subdominant at all times compared to the buoyancy and viscous ones: the amplitudes of $D(t)$\footnote{Notice that $D(t)$, which is the hyper-viscous version of what in the literature is called ``helical enstrophy'' \cite{rorai2013helicity} or ``superhelicity'' \cite{hide1989superhelicity}, is a pseudoscalar like helicity, and they are both sign-indefinite.} and $F(t)$ are, respectively, two and one order of magnitude smaller than those of $B(t)$ and $V(t)$. 
A comparison with Fig.~\ref{fig:helicity_vs_time_and_pdf}(a) reveals that the surge in helicity -- as well as its later decay back to zero -- is the result of the constructive contribution of buoyancy and viscous terms: 
when their sum is positive (around $t\simeq40\, \tau_{NL}$) they produce a net increase of helicity, and when negative (period from $t\simeq200\, \tau_{NL}$ to $t\simeq1500\, \tau_{NL}$) they decrease it towards negative values. 

Finally we discuss the connection between helicity dissipation and interface migration. The dynamics of the density field $\phi$--profile, derived by applying a plane-average to eq.~\eqref{eq:phi_evol}, is governed by an advection-diffusion equation:
\begin{equation}
    \partial_t \langle \phi \rangle = - \partial_z \langle u_z \phi \rangle -\kappa (-\partial_{zz})^2 \langle \phi \rangle\ ,
    \label{eq:phi_profile_evol}
\end{equation}
where the (hyper-)diffusive term may be neglected in the present reasoning. 
On the other hand we saw that helicity is produced or destroyed by buoyancy according to: $\partial_t h \simeq 2 B$.
If a flow is close to being fully helical -- locally in specific planes or globally in the whole volume --
then the relation
\begin{equation}
    \boldsymbol{\omega} \propto \pm\, \boldsymbol{u}
    \label{eq:omegabetau}
\end{equation}
holds, up to small corrections. 
{\color{black} As we discuss in appendix B (see also Fig.~\ref{fig:helicitybalance_volumeavg_vs_time}), observations suggest that migration of interfaces occurs at the locations where helicity undergoes dissipation, i.e. $\partial_t |h| < 0$.
Taking plane averages, the condition of helicity decay by buoyancy is $\langle B\ \text{sign}(h) \rangle < 0$, which, upon using \eqref{eq:omegabetau}, corresponds to the condition that the vertical flux of $\phi$ satisfies $\langle u_z \phi \rangle > 0$. 
Therefore, helicity dissipation is directly linked to vertical migration of the density structure and to interface merging. }
\textcolor{black}{
The evolution of the ramp-cliff structure in Fig.~\ref{fig:interfacemerging_SofiaN18} is the manifestation of such effect: if $\langle u_z \phi \rangle>0$ when two interfaces are close, then the lower one is pushed upward by the advancing layer, while the upper one moves downward for the same reason. 
We refer to appendix B for a more detailed explanation, and a discussion of mean helicity balance and the spatiotemporal evolution of $\langle B \rangle$.\\
Summarising, the sign of $\langle B \rangle + \langle V \rangle$ determines the chirality of the emerging helical structures, while 
the convergence of two close interfaces, containing a thin layer of fluid in between, is promoted by the fluid regions in which buoyancy-induced helicity dissipation occurs.}


\section{Discussion and perspectives}

We have presented a characterization of the coarsening dynamics of the vertical density profile in forced, stably stratified turbulent flows, with a focus on the process of interface merging. 
Alongside interface decay, these are the two processes -- observed in experiments and simulations -- leading any step-like density profile to its asymptotic equilibrium state. 
Building on a previous study \cite{cocciaglia2025longtime}, we carried out simulations with different $Fr$, showing interface decay at $Fr=0.22$ and interface merging at $Fr=0.076$. 
The kinetic energy excursions observed at lower $Fr$, occurring synchronously with enhanced vertical drift of interfaces, receive the strongest contribution from the shear component (purely horizontal velocity with no variability along planes), while shear energy returns subdominant with respect to wave and vortical ones during stationary periods outside the excursions. 
Shear-energy spectra show very steep, $Fr$-independent slopes below the forcing scale, so the small-scale behavior is governed by interacting waves and vortices.

We observed that the horizontal, shear-dominated flow takes the shape of frozen circularly-polarized waves, representing a fully-helical ABC-like flow. The spontaneous emergence of chirality is explained studying the terms involved in the helicity balance equation. 
The helicity variations are mainly caused by the large scale buoyancy production term $B(t)$, which can be connected to the vertical buoyancy flux for helical mean flows, and the small-scale viscous term $V(t)$ (the latter being known as helical enstrophy or superhelicity \cite{rorai2013helicity, hide1989superhelicity}).  On the basis of these observations, we suggest that vertical movement of the density layering structure, and especially the interface merging phenomenon, is associated to -- and arguably caused by -- large non-chiral structures in the fluid velocity. 

While helicity is spontaneously created when in rotating flows  \cite{marino2013emergence,gome2026}, the emergence of mean helicity in non-rotating stably-stratified turbulence is a topic deserving further investigation, which will be addressed in future studies.
Despite rotating, convective dynamics explicitly promoting helicity production \cite{lilly1986helicity}, non-forced simulations of Boussinesq flows showed instead that a random initial condition with zero helicity do not show mirror-symmetry breaking \cite{rorai2013helicity}. 
It is also interesting to note that hyper-dissipative effects, that are also employed in simulations of active fluids, are shown to cause the emergence of large helical structures \cite{slomka2017spontaneous}. 
Additional studies, in terms of e.g. transfer among helical modes \cite{biferale1998helicity,sahoo2017}, can help elucidate these issues.


\acknowledgments
We acknowledge EuroHPC Regular Access, project No. EHPC-REG-2021R0049, for the computational resources.
This work was supported by the European Research Council under the European Union’s Horizon 2020 research and innovation programme Smart-TURB (ERC Grant Agreement No. 882340). 
N.C. acknowledges support from the National Recovery and Resilience Plan (PNRR), Mission 4 Component 2 Investment 1.1 - Call No. 104 – Project: CO-SEARCH, code 202249Z89M – CUP E53D23001610006, 
and from Fondo Italiano per la Scienza 2022-2023 (FIS2), CUP E53C24003760001. 
A.S.L. acknowledges financial support from the NextGeneration EU plan and MUR, call PRIN2022PNRR - Project: BREAKUP, P20225AEF4 — CUP B53D23028750001.

\bibliographystyle{unsrt}
\bibliography{bib}

\end{document}